\documentclass[a4paper,11pt]{article}
\pdfoutput=1 

\usepackage{jheppub}
\usepackage[T1]{fontenc} 
\usepackage{amsmath,amssymb,amsthm}
\usepackage{latexsym,graphicx,bbm,bookmark}

\newcommand{\beq}{\begin{eqnarray}}
\newcommand{\eeq}{\end{eqnarray}}

\newcommand{\D}{\mathcal{D}}
\newcommand{\p}{\partial}
\newcommand{\Tr}{{\rm Tr}}

\title{\boldmath Mass deformed world-sheet action of  semi-local vortices}

\author[a,b]{Yunguo Jiang}

\affiliation[a]{School of Space Science and Physics, \\ Shandong University at Weihai,  264209 Weihai, China}
\affiliation[b]{Shandong Provincial Key Laboratory of Optical Astronomy \\ and Solar-Terrestrial Environment, 264209 Weihai, China}

\emailAdd{jiangyg@sdu.edu.cn}

\abstract{The mass deformed effective world-sheet theory of semi local vortices was constructed via the field theoretical method. By Euler-Lagrangian equations, the Ansatze for both the gauge field and the adjoint scalar were solved, this ensures that zero modes of vortices are  minimal excitations
of the system.  Up to the $1/g^2$ order, all profiles are solved. The mass deformed effective action was obtained by integrating out the transverse plane of the vortex string. The effective theory interpolates between the local vortex and the lump. Respecting certain normalization conditions, the effective theory shows a Seiberg-like duality, which agrees with the result of the K\"ahler quotient construction.}
\arxivnumber{xxxxxxxx}

\begin{document}

\newpage
\pagenumbering{arabic}
\setcounter{page}{1}
\setcounter{footnote}{0}
\renewcommand{\thefootnote}{\arabic{footnote}}
\maketitle
\section{Introduction}

The non-Abelian vortices are considered as an important tool to understand the non-Abelian confinement in ${\cal N}=2$ supersymmetric QCD \cite{Auzzi:2003fs,Hanany:2003hp,Shifman:2004dr}. The low energy effective theory of non-Abelian vortices plays an important role in understanding how monopoles are confined on the vortex string and why spectra of the two and four dimensional theories coincide. The construction of the effective action started as soon as the discovery of the non-Abelian vortices, and has been extensively investigated \cite{Hanany:2004ea,Shifman:2006kd,Gorsky:2004ad,Shifman:2011xc,Eto:2007yv,Gudnason:2010rm,Eto:2011cv}.

Hanany and Tong used the index theorem to obtain the dimension of the moduli space of  $U(N)$ non-Abelian vortices, and studied the low energy dynamics of vortex strings for both local and semi-local cases \cite{Hanany:2003hp,Hanany:2004ea}. The spectra of the vortex string coincide with that of the four dimensional parent supersymmetric gauge theory, which is a proof of Dorey's  2d-4d duality \cite{Dorey:1999zk}. Shifman et al derived the world-sheet theory of semi local non-Abelian strings by a field theoretical method \cite{Shifman:2006kd,Shifman:2011xc}. For semi-local vortices, the transverse size of the magnetic flux is not fixed but becomes a modulus. By introducing an infrared regulator, a so-called $zn$ model was obtained for the single semi-local vortex with or without twisted mass \cite{Shifman:2011xc}. Using the moduli matrix method, Eto et al. studied the moduli space of high winding semi-local vortices, and found that dynamical variables in the effective action, including  orientational zero modes and size moduli,  depend on the point of the moduli space \cite{Eto:2007yv}. Recently, the field theoretical method was generalized to construct the world-sheet action of fundamental $SO$ and $USp$ vortices, and applied to some high winding vortices in $U(N)$ and $SO(2N)$ theories \cite{Gudnason:2010rm}. Further, Eto et al. derived the mass deformed sigma models, and showed that confined monopoles are kinks on the vortex string for $SO$ and $USp$ theories  \cite{Eto:2011cv}.

The aim of this paper is to derive the mass deformed effective action of  semi local $U(N)$ vortices. There are two alternative ways to construct the effective action of vortices. One is the moduli matrix formalism \cite{Eto:2006cx,Eto:2007yv,Eto:2011cv,Eto:2012qda,Cipriani:2012pa}, and another one is the field theoretical method \cite{Shifman:2006kd,Gorsky:2004ad,Shifman:2011xc,Gudnason:2010rm}. Besides that, the D-brane construction is also a powerful technique to obtain the effective potential on the vortex moduli space \cite{Hanany:2004ea}. In the moduli matrix formula, all zero modes are encoded in  components of the matrix representation of fields, and  moduli matrices are holomorphic with respect to a complex variable. The moduli space  and the effective action have been analyzed and obtained by this method \cite{Eto:2006cx,Eto:2007yv,Cipriani:2012pa}. In the field-theoretical method, the Ansatz of the field is composed of  radial profiles and  reducing matrices.  The two dimensional effective action composed of  reducing matrices can be obtained by integrating out  profile functions from the bulk four dimensional theory \cite{Gorsky:2004ad,Gudnason:2010rm}. So, we need to know the representation of the reducing matrix for the corresponding vortex configuration. However, the known reducing matrices are quite limited \cite{Delduc:1984sz}, this hinders the application of the field theoretical method. There are by-products of this method, i.e., the analytical solutions for the profiles of the gauge fields and the adjoint scalars. For instance, the profile for the gauge fields in the time and string directions has been solved for the  $k=2$ $SO(2N)$ vortex \cite{Gudnason:2010rm}. In the mass deformed case, the profile for the adjoint scalar was also obtained in the same way \cite{Eto:2011cv}. In this paper, we choose the field theoretical method to work out the mass deformed effective theory for semi-local non-Abelian  vortices. For $U(N)$ vortices, the Ansatz was invented by minimizing the massive excitations \cite{Shifman:2006kd,Gorsky:2004ad}. The systematic method to obtain the right Ansatze is to solve the Euler-Lagrangian (EL) equation, and the system reserves with the low energy excitations in such a way \cite{Eto:2012qda}. When constructing the effective action of  semi local vortices, the Anzatz for the adjoint scalar was not studied  \cite{Shifman:2011xc}. Until very recently, Bolokhov et al. investigated the Anzatz of the adjoint scalar for the local vortices \cite{Bolokhov:2013bea}. We will solve  EL equations to obtain the Ansatze for both the gauge field components and the adjoint scalar.

This paper is organized as follows. In Section \ref{sec:model}, we describe the model, and give the semi local vortex solutions. In Section \ref{sec:ans}, we use  EL equations to obtain Ansatze for the gauge field and the adjoint scalar field, respectively. On the basis of these Ansatze, we constructed the mass deformed effective action of semi local vortices. We also discussed the effective theory of high winding vortices  in Section \ref{sec:efa}. Discussions and conclusions are given in the last section.

\section{The model \label{sec:model}}

The bulk theory is the ${\cal N}=2$ supersymmetric QCD with $U(N_c)$ gauge  and $SU(N_f)$ flavor group transforming in the fundamental representation. The $N_c=N_f$ case will describe the local non-Abelian vortex solutions \cite{Auzzi:2003fs,Hanany:2003hp,Shifman:2004dr}, but we restrict to the  $U_f > N_c$ case, which will describe the non-Abelian semi local vortices \cite{Shifman:2006kd,Eto:2007yv}. Besides  fermions, the theory contains a $U(N_c)$ vector multiplet $A_{\mu}$, an adjoint scalar  $\Phi$, and two chiral multiplet $q$  and $\tilde{q}$. The theory also has a Fayet-Illiopoulos parameter $\xi>0$, which forces the theory onto the Higgs branch. We also set that the anti-fundamental  multiplet $\tilde{q}$ is zero, otherwise, there are no BPS vortex solutions.   With these setups, the bosonic truncation of the Lagrangian is written as follows  \cite{Shifman:2011xc}
\begin{align}
{\cal L}_{4d} = \Tr \bigg\{ &-\frac{1}{2g^2} F_{\mu \nu} F^{\mu \nu}
 + \D_{\mu} q (\D^{\mu}q)^{\dag}+ \frac{1}{g^2}(\D_{\mu}\Phi)\D^{\mu} \Phi \nonumber \\
 &-\frac{g^2}{4}(q q^{\dag}-\xi\mathbf{1}_{N_c})^2-|\Phi q+ q{\rm \bf M}|^2
 \bigg\}.  \label{eq:model}
 \end{align}
The squark field $q$  is  written as an $N_c\times N_f$ matrix, and the adjoint scalar $\Phi$ is of an $N_c\times N_c$ matrix. The expression of the mass matrix ${\rm \bf M}$ is generic, namely, ${\rm \bf M}={\rm diag} (m_1, \ldots, m_{N_f})$,  which breaks the flavor group $SU(N_f)$ down to $U(1)^{N_f-1}$, if all  masses are non-degenerate.
The gauge couplings of the Abelian and non-Abelian components are set to be equal for simplicity.
In convention, the covariant derivatives and the gauge field tensor are written as follows
\begin{align} \label{eq:covder}
\D_{\mu}q=&\p_{\mu}q+i A_{\mu}q, \\
  \D_{\mu} \Phi=&\p_{\mu}\Phi+i[A_{\mu},\Phi], \\
  F_{\mu \nu}=& \p_{\mu} A_{\nu}-\p_{\nu} A_{\mu}+i[A_{\mu}, A_{\nu}].
\end{align}
Note that the gauge field $A_{\mu}$ contains both the Abelian and non-Abelian components, i.e., $A_{\mu}=A_{\mu}^0 t^0+ A_{\mu}^a t^a$. The normalization is taken to be $\Tr (t^a t^b)= 1/2 \delta^{ab}$ and $t^0 \equiv \mathbf{1}_{N_c}/\sqrt{2N_c}$. We choose to work in such a vacuum that the mass matrix takes the form ${\rm \bf M}={\rm diag} (m_1, \ldots, m_{N_c}, m_{N_c+1}, \ldots, m_{N_f})$ and the first $N_c$ flavors of the squark $q$ are condensed,
\beq \langle \Phi \rangle =-{\rm diag}(m_1, \ldots, m_{N_c}), \qquad \langle q \rangle =\sqrt{\xi} (\mathbf{1}_{N_c}, \,
0_{\tilde{N}}), \label{eq:vacuum} \eeq
where $\tilde{N}=N_f-N_c$. The reduced Higgs branch of the vacua is the Grassmannian, ${\it Gr}_{N_c, N_f}$. In the strong coupling limit, the moduli space of the semi-local vortex ${\cal V}_{k,(N_c, \tilde{N})}$ becomes to the Moduli space of Grassmannian ${\it Gr}_{N_c, \tilde{N}}$ lump.

The mass parameters $m_i$ are tuned to  a common value $m_i \approx m$, satisfying the constraint $m \ll \sqrt{\xi}$. In the  regime $\Lambda \ll \sqrt{\xi}$, a color-flavor locking symmetry $SU(N_c)_{\rm diag}$ remains, which develops  vortex configurations. Therefore, the theory experiences a hierarchy symmetry breaking, i.e.,
\beq \label{eq:sbp}
U(N_c) \times SU(N_f) \mathop{\longrightarrow}^{\sqrt{\xi}} S[U(N_c)_{\rm diag} \times U(\tilde{N})_f] \mathop{\longrightarrow}^{m} H_{c+f} \times \tilde{H},
\eeq
where $H_{c+f}$ is a subgroup of $SU(N_c)_{\rm diag}$ depending on the setting of $m_i$, and $\tilde{H}$ denotes a remaining global flavor symmetry group. The generic $m_i$ will break the color-flavor symmetry. However, we assume that such breaking is in a very weak manner. This will produce a ``shallow'' potential for the world-sheet action of  vortices.
When $\tilde{N}$ becomes zero, the system reduces to the local non-Abelian vortex case, where a narrow size flux tube confines monopoles as kinks on the string\cite{Shifman:2006kd,Eto:2011cv}. When $\tilde{N} \neq 0$, we can shed some light on how monopoles are confined by the semi-local vortex.

\subsection{The moduli space}

According to the index theorem, the dimension of the moduli space ${\cal V}_{k, (N_c, \tilde{N})}$ is $2k(N_c+ \tilde{N})$ \cite{Hanany:2003hp}.
The moduli space of the vortex configuration can be constructed by the K\"ahler quotient method. Following \cite{Eto:2007yv},we use the moduli matrix formalism to show it in detail. The squark $q(z)$, whose elements are polynomials in z and the corresponding coefficients are coordinates on the moduli space, can be written as follows
\begin{equation}\label{eq:qz}
  q(z)=(D(z),Q(z)),
\end{equation}
 where $D(z)$ and $Q(z)$ are $N_c\times N_c$ and $N_c \times \tilde{N}$ matrices, respectively. For the winding number $k$, $q(z)$ has the degree of $k$. $q(z)$ indicates the vortex configuration evidently.
By  proper relations (see section 2 in \cite{Eto:2007yv}), all  moduli coordinates can be collected in the set of  constant matrices $\big( \mathbf{Z},\mathbf{\Psi},\tilde{\mathbf{\Psi}}\big)$ modulo the $GL(k,\mathbf{C})$.  $\mathbf{Z}$, $\mathbf{\Psi}$, and $\tilde{\mathbf{\Psi}}$ are
constant $k\times k$, $N_c\times k$ and $k \times \tilde{N}$ matrices, respectively. By counting  dimensions, one can easily verify that the triplet indeed represent moduli space coordinates, although the quotient space has the non-Hausdorff properties. Given a vortex configuration $q(z)$, one can obtain the expression of the triplet uniquely after fixing the $GL(k,\mathbf{C})$ action.

According to their physical characteristic,  zero modes of  semi-local vortices can be classified into tree types, namely the positional, the orientational and the size moduli. In addition, the moduli can be further classified into the normalizable and non-normalizable categories. The number of nonmalizable zero modes (NZMs) is subtle. Let us define that $r \equiv {\rm rank} (\mathbf{\Psi} \tilde{\mathbf{\Psi}})$ and $j \equiv {\rm min}(k, N_c, \tilde{N})$. For the case of $k \leq {\rm min}(N_c, \tilde{N})$, the number of NZMs  is $2k^2$. For the case of $k \geq {\rm min}(N_c, \tilde{N})$, the number of NZMs is $2(kN_f-N_c \tilde{N})$. When $r<j$, it was called that the NZMs are enhanced \cite{Eto:2007yv}. 

In the construction of the effective action, these non NZMs must be fixed, since they are not dynamical. While the rest moduli are allowed in the geodesic approximation. Here, we present the moduli space in the moduli matrix formula, which indeed has the corresponding reducing matrix formula \cite{Eto:2006cx,Gudnason:2010rm}. For constructing the mass deformed effective action of  semi local vortices, we choose to work in the reducing matrix formula, and use the Ansatz given in Ref. \cite{Shifman:2011xc}.
As stated above, the dynamics of vortices depends on the point of the moduli space on which we work. For the high winding case, we choose a special point, namely the co-axial vortices, to construct the effective action as a concrete example. However, this is not a generic point in the moduli space. We will leave the construction of the effective action on a generic point for the future work.

\subsection{The vortex solution}

Consider the static configurations of the model in Eq.(\ref{eq:model}), and suppose that the vortex string lies along the $x_3$ direction.
The Lagrangian which has vortex solutions is written as follows
\beq \label{eq:Lag0}
{\cal L}^{0}_{\rm vortex}=\Tr \bigg\{ -\frac{1}{g^2} F_{12} F^{12}
 + \D_{i} q (\D^{i}q)^{\dag} -\frac{g^2}{4}(q q^{\dag}-\xi\mathbf{1}_{N})^2 \bigg\}.
\eeq
Here $i=1,2$ denotes the directions in the transverse plane of vortices. Other terms in Eq.(\ref{eq:model}) are ignored at this moment, they are also important for the construction of effective theory. After the Bogomol'nyi completion, the system has a bound energy.  The Bogomol¡¯nyi-Prasad-Sommerfield (BPS) equations are written as follows
\begin{align}
&(\D_{1}+i \D_2)q=0, \\
&F_{12}-\frac{g^2}{2}(q q^{\dag}-\xi\mathbf{1}_{N})=0.
\end{align}
 For simplicity, we choose $\tilde{N}=1$ at first, which means there is only one ``additional flavor''.
Consider a high winding vortex which has the configuration as follow,
\beq \label{eq:sq}
q= \begin{pmatrix}
             \phi_1(r)e^{ik\theta}& & & &\phi_3(r)\\
              &\phi_2(r)&& &\\
             &&\ddots& &\vdots\\
            &&& \phi_2(r) & 0
          \end{pmatrix}.
\eeq
Notice that the winding term appears at the first diagonal component of the matrix, which describes that $k$ fundamental vortices have the same positional moduli, namely the co-axial vortices. In this way, the moduli space of the vortex are highly reduced.  In the additional flavor, one has a profile $\phi_3(r)$, which contains the size moduli \cite{Vachaspati:1991dz,Hindmarsh:1991jq,Shifman:2011xc}. The orientational moduli can be tuned on by the color-flavor rotation, which we will discuss in the following.
Requiring $D_i q |_{\rm r \to \infty} \to 0$, the gauge fields $A_i$ is given as follows
\beq
A_i=\epsilon_{ij}\frac{x_j}{r^2} \begin{pmatrix}
k-f(r)& & & \\
              &0&& \\
             &&\ddots& \\
            &&&  0
          \end{pmatrix}. \eeq
By going to the singular gauge, i.e., $q \to U q$, $A_i \to U A_i U^{\dag} +{\rm i} \p_i U U^{\dag}$,
the $q$ and $A_i$ can be rewritten as follows
\beq \label{eq:q0} q= \begin{pmatrix}
             \phi_1(r)& & & &\phi_3(r)e^{-ik\theta}\\
              &\phi_2(r)&& &\\
             &&\ddots& &\vdots\\
            &&& \phi_2(r) & 0
          \end{pmatrix}, \qquad A_i= -\epsilon_{ij}\frac{x_j}{r^2}f(r) \begin{pmatrix}
1& & & \\
              &0&& \\
             &&\ddots& \\
            &&&  0
          \end{pmatrix}.
\eeq
Here $U={\rm diag}(e^{-ik\theta}, 1, \cdots, 1)$. The merit of the singular gauge is that  BPS equations are easily to be solved.

With these configurations,  BPS equations for profiles are given by
\begin{align}
\phi_1'(r)-\frac{f(r)}{r} \phi_1(r)=&0, \label{eq:bps1} \\
\phi_2^2(r)=\xi, \qquad \phi_2'(r)=&0,  \label{eq:bps2}\\
\phi_3'(r)+\frac{k-f(r)}{r} \phi_3(r)=&0,  \label{eq:bps3}\\
\frac{1}{r}f'(r)-\frac{g^2}{2}[\phi_1^2(r)+\phi_3^2(r)-\xi]=&0. \label{eq:bps4}
\end{align}
Here the prime sign denotes the derivative with respect to $r$. At the infinity,  boundary conditions are given by
\beq  \phi_1(\infty)=\sqrt{\xi}, \qquad \phi_2(\infty)=\sqrt{\xi},\qquad
\phi_3(\infty)=0, \qquad f(\infty)=0.
\eeq
When $r$ goes to zero, boundary conditions are written as follows
\beq \label{eq:bcr0}
\phi_1(0)=0, \qquad \p_r \phi_2(0)=0, \qquad f(0)=k.
\eeq
The solution for $\phi_2$ in Eq.(\ref{eq:bps2}) can be solved directly,
\beq \phi_2(r)= \sqrt{\xi}, \eeq
which means that flavors from $2$ to $N_c$ have trivial profiles. We also notice that if
\beq \phi_3(r)= \frac{\rho}{r^k} \phi_1(r), \label{eq:phi3} \eeq
then Eq.(\ref{eq:bps1}) and Eq.(\ref{eq:bps3}) coincide, where $\rho$ is  the size modulus of
the semi local vortex. $|\rho|$ describes the size of the vortex, while there is still a $U(1)$ phase rotation freedom for $\rho$.

It is natural to ask if there are other profiles for the additional $\tilde{N}=1$ flavor. For  high winding semi local vortices,
it is indeed the case \cite{Leese:1992fn}. For example, the profile of the $\tilde{N}$ flavor can be written as $\phi_3=\sum_{n=1}^{k-1}\varphi_n e^{i(k-n)\theta}$.  There are more profile equations and boundary conditions for $\varphi_n$, which are difficult to solve. The difficulty here originates from the disadvantage of the reducing matrix method for high winding vortices. However, we find it solvable in the strong coupling limit if we stick to the formula in Eq.(\ref{eq:phi3}).

With such setting, we have only one size modulus in the moduli space.
The  remaining two independent equations are written as follows
\begin{align}
\phi_1'-\frac{f}{r} \phi_1=&0, \label{eq:bpsi1} \\
\frac{1}{r}f'-\frac{g^2}{2}[\phi_1^2(1+\frac{|\rho|^2}{r^{2k}})-\xi]=&0. \label{eq:bpsi2}
\end{align}
In the large gauge coupling $g \to \infty$, these two equations can be solved
algebraically in the $1/g^2$ expansion. Keeping only terms of the order of
 $1/g^2$,  profile functions can be written as follows
\beq \phi_1= \phi_{1,0} + \frac{1}{g^2} \delta \phi_1, \qquad f= f_0 +
\frac{1}{g^2} \delta f.  \label{eq:expansion} \eeq
 Substituting $\phi_2=\sqrt{\xi}$ into Eqs. (\ref{eq:bpsi1}) and (\ref{eq:bps2}), the zeroth order equations are written
 as follows
\begin{align}
\phi_{1,0}^2(1+\frac{|\rho|^2}{r^{2k}})-\xi=&0, \\
\phi_{1,0}'-\frac{f_0}{r} \phi_{1,0}=&0.
\end{align}
And equations with the order of $1/g^2$ can be expressed as
\begin{align}
\delta \phi_1'-\frac{1}{r}(\phi_{1,0}\delta f + f_0 \delta \phi_1)=&0, \\
\frac{1}{r}f_0' -  \phi_{1,0} \delta \phi_1 (1+\frac{|\rho|^2}{r^{2k}})=&0.
\end{align}
After some algebra calculations, we obtain solutions as the following
\begin{align}
\phi_{1,0}&=\sqrt{\xi}\frac{r^k}{\sqrt{r^{2k}+|\rho|^2}}, \label{eq:phi1} \\
f_0&=k\frac{|\rho|^2}{r^{2k}+|\rho|^2} \label{eq:phi1f},    \\
 \delta \phi_1&=- \frac{2k}{\sqrt{\xi}} \frac{|\rho|^2
  r^{3k-2}}{(r^{2k}+|\rho|^2)^{\frac{5}{2}}}, \\
   \delta f&=\frac{4k}{\xi}
\frac{|\rho|^2 r^{2k-2}}{(r^{2k}+|\rho|^2)^3} \bigg[k(r^{2k}-|\rho|^2)+r^{2k}+|\rho|^2\bigg].
\end{align}
Since $k$ is a positive integer,  boundary conditions for $r \to  \infty$ are satisfied. When $k$=1, the results
agree with Shifman et al. \cite{Shifman:2011xc}.

Degenerate vortex solutions can indeed be generated by color-flavor  rotations. For $\tilde{N}=1$, there is a remaining global flavor group, i.e.,
$\tilde{H}=U(1)_f$ in Eq.(\ref{eq:sbp}). However, this $U(1)$ rotation can be absorbed into the phase transition of size modulus $\rho$. The global color-flavor group $SU(N_c)$ is broken by the vortex configuration in Eq.(\ref{eq:sq}). The symmetry breaking pattern produces the ``Nambu-Goldstone'' modes $\mathbbm{C}P^{N-1} \cong SU(N_c) /SU(N_c-1)\times U(1)$, which are represented by  the reducing matrix $U$.
 Now, we turn on the color-flavor transformations here, i.e., $q \to U\,q \, \tilde{U}$ and $A_i \to U A_i \,U^{\dag}$, where $U$ and $\tilde{U}$ are written as follows \cite{Gudnason:2010rm}
\begin{align}
U =\begin{pmatrix}
X^{-\frac{1}{2}} & - B^\dag Y^{-\frac{1}{2}} \\
B X^{-\frac{1}{2}} & Y^{-\frac{1}{2}}
\end{pmatrix}, \qquad \tilde{U}=\begin{pmatrix}
U^{\dag} &  \\
 & 1
\end{pmatrix}.
\label{eq:Umatrix}
\end{align}
Note that $\tilde{U}$ is composed of $U^{\dag}$ and $1$ as the diagonal components. The elements  $X$ and $Y$ in the reducing matrix $U$ are defined by
\beq
X\equiv 1 + B^\dag B \ , \quad
Y\equiv\mathbf{1}_{N_c-1} + B B^\dag \ ,
\eeq
where $B$ is an $N-1$ component column vector. Its relation with the vector $n$ in Ref. \cite{Gorsky:2004ad,Shifman:2011xc} is expressed as
\beq
 n=\begin{pmatrix}
 X^{-1/2}\\
 BX^{-1/2} \end{pmatrix}.
\eeq
From now on, we use the vector $n$ and $n^{\dag}$ to represent  modulus parameters. The squark $q$ and the gauge fields $A_i$ are written as follows
\begin{align}
q(r,x_{\alpha}) &= \left( \phi_2 \mathbbm{1}_{N_c}+(\phi_1-\phi_2) \, n n^{\dag} \,\,\vline \,\, \phi_3 e^{-ik\theta} n \right), \label{eq:q}\\
A_i(r,x_{\alpha}) &= -\epsilon_{ij}\frac{x^j}{r^2} f(r) \,\, n n^{\dag},
  \label{eq:Ai}
\end{align}
where $x_{\alpha}=(x_0,x_3)$. Note that $nn^{\dag}$ is an $N_c \times N_c$ matrix.
 The zero modes are not genus zero-modes of the system, they are massive modes in the vacuum, but massless along the vortex string. Thus, they can propagate along the vortex string direction in the vicinity of the axis.
 We promote the collective modes $n$ and $\rho$ to be dependent of $x_{\alpha}$, i.e.,
\beq n \to n(x_{\alpha}), \qquad \rho \to \rho(x_{\alpha}).  \eeq
The collective coordinates of the moduli space are considered to be the fluctuation fields
around the background solution.  Also there is one positional mode which stands for the mass center of the
vortex,  it's dynamics is straightforward, we will not discuss them in the following.

\section{The Ansatze \label{sec:ans}}

Generically, the motion of  solitons can be well approximated by the geodesic motion in the moduli space \cite{Manton:1981mp}.
In the BPS configuration, vortices are static. The $x_{\alpha}$-dependence of the moduli parameters induces small deviations from the background BPS configuration. The momentum of the motion of  zero modes is considered to a small variable for the slow moving case. A systematic method to construct the effective action and its high derivative corrections of  BPS solitons in  supersymmetry gauge theories has been presented in the moduli matrix formalism \cite{Eto:2006pg,Eto:2006uw,Sakai:2007fm,Eto:2012qda}. Cipriani and Fujimori used this formalism to study the effective action of vortex-monopole complex \cite{Cipriani:2012pa}. Here, we take the spirit of this method and apply it to the field theoretical formalism. We also assume that the excitation energy of  zero modes is much less than the typical mass scale of  massive modes, i.e., $\p_{\alpha} \ll g\sqrt{\xi}$.

The Lagrangian can be expanded with respect to the derivative $\p_{\alpha}$. The zeroth order terms give the static vortex configuration, and becomes the bounded energy after the Bogomolnyi completion. The lowest order terms take the form of  two derivatives for  zero modes, i.e., the non linear sigma models with certain target space. The second order terms  belong to high derivative corrections ${\cal O}(\p_{\alpha}^4)$, which are not considered in this paper.
The Lagrangian in Eq.(\ref{eq:model}) can be decomposed as
\begin{align}
{\cal L}_{\rm 4d} &= {\cal L}^{0}_{\rm vortex}+{\cal L}^{2}_{\rm eff}+{\cal O}(\p_{\alpha}^4), \\
{\cal L}^{2}_{\rm eff}&= \Tr \bigg\{ -\frac{1}{g^2} F_{i \alpha} F^{i \alpha}+ \frac{1}{g^2}|\D_{i}\Phi|^2+\D_{\alpha} q (\D^{\alpha}q)^{\dag}-|\Phi q+ q{\rm \bf M}|^2
.    \bigg\}, \label{eq:Lag1}
\end{align}
where ${\cal L}^{0}_{\rm vortex}$ has already been given in Eq.(\ref{eq:Lag0}). The adjoint
scalar $\Phi$ has the order of $\p_{\alpha}$, and contains fluctuations of collective modes.

In the field theoretical method, the effective theory is obtained by integrating out the $x_i$ ($i=1,2$) coordinates for the $4$ dimensional bulk action. We still have to give the unknown Ansatze of $A_{\alpha}$ and $\Phi$, which are unknown. Shifman et al. have constructed the Ansatz of $A_{\alpha}$ for the $U(N_c)$ local vortex in Refs.  \cite{Gorsky:2004ad, Shifman:2004dr,Shifman:2006kd} and for the semi local $U(N_c)$ vortex in the Ref. \cite{Shifman:2011xc}. Gudnason et al. have given the Ansatz of $A_{\alpha}$, including the $SO$ and $USp$ cases with the help of reducing matrices \cite{Gudnason:2010rm}. The idea there is to design this ``input parameter'' to minimize the action. However, in the moduli matrix formalism,  the Ansatze of both $A_{\alpha}$ and $\Phi$ were solved by equations of motion \cite{Eto:2006pg,Eto:2011cv,Cipriani:2012pa}. Inspired by this, we will use equations of motion for $A_{\alpha}$ and $\Phi$ to obtain their Ansatze. This will ensure that the formula of them indeed minimizes the action.

\subsection{The gauge field}

The second order Lagrangian has been given in Eq.(\ref{eq:Lag1}). The equation of motion for $A_{\alpha}$ is calculated to be
\begin{align} \label{eq:Aalpha}
0= &  \frac{1}{g^2} {\cal D}^i F_{i\alpha}^{a} -i  \Tr \big[ t^{a} ({\cal D}_{\alpha}
q) q^{\dag} - t^{a} q({\cal D}_{\alpha}q)^{\dag} \big],
\end{align}
where $a=0,1, \cdots, N_c^2-1$ is the generator index of the $SU(N_c)$ group.
With the vortex configuration, $q$ and $A_{i}$ have already been given in Eqs.(\ref{eq:q}) and (\ref{eq:Ai}), respectively.
Substituting them into  Eq.(\ref{eq:Aalpha}), the equation of motion for $A_{\alpha}$ is given by
\begin{align}
0=& \frac{1}{r}\p_r(r\p_r)A_{\alpha}+i \frac{f^2}{r^2}
[(\p_{\alpha}n)n^{\dag}-n\p_{\alpha}n^{\dag}+2(\p_{\alpha} n^{\dag} n)
\,nn^{\dag}]-\frac{f^2}{r^2}[\{A_{\alpha}, nn^{\dag}\}-2 nn^{\dag}A_{\alpha}nn^{\dag}] \nonumber \\
&-\frac{g^2}{2}\bigg( 2\phi_2^2A_{\alpha}+(\phi_1^2+\phi_3^2-\phi_2^2)\{A_{\alpha}, nn^{\dag}
\}-i[\p_{\alpha}(\phi_3 n) (\phi_3 n)^{\dag}- (\phi_3 n )\p_{\alpha} ( \phi_3 n)^{\dag}]  \nonumber \\
&-i(\phi_1-\phi_2)^2[\p_{\alpha}nn^{\dag}-n\p_{\alpha}n^{\dag}+2(\p_{\alpha} n^{\dag}
n)\,nn^{\dag})] \bigg).
\end{align}
Note that $n$ is a normalized vector with $N_c$ components, which satisfies the condition $n^{\dag}n=1$.

In order to solve this equation, properties of $A_{\alpha}$ should be discussed. First, $A_{\alpha}$ is Hermitian. Secondly, we require that $A_{\alpha}$ can reduce to the formula for  local vortices. We use the separation of variables method to express $A_{\alpha}$ as a product of the profile and the matrix containing zero modes. Based on these arguments,
 we suppose that $A_{\alpha}$  takes such a formula
\beq   A_{\alpha}= \omega(r) W_{\alpha}+ \gamma(r) \tilde{W}_{\alpha}, \label{eq:Asetup} \eeq
where $\omega(r)$ and $\gamma(r)$ are two profiles with respect to only one variable $r$, while $W_{\alpha}$ and $\tilde{W}_{\alpha}$ are two matrices describing orientational zero modes. Both $W_{\alpha}$ and $\tilde{W}_{\alpha}$ are Hermitian. 

Substituting the formula of $A_{\alpha}$ into the  equations of motion, one obtains
\begin{align} \label{eq:tem}
0=& \frac{1}{r}\p_r(r\p_r)\omega W_{\alpha}+ \frac{f^2}{r^2}
i[(\p_{\alpha}n)n^{\dag}-n\p_{\alpha}n^{\dag}+2(\p_{\alpha} n^{\dag} n)
\,nn^{\dag}] \nonumber \\
&-\omega\frac{f^2}{r^2}[\{ W_{\alpha}, nn^{\dag}\}-2 nn^{\dag} W_{\alpha}nn^{\dag}] -\frac{g^2}{2}\bigg( (\phi_1^2+|\phi_3|^2-\phi_2^2)\omega \{W_{\alpha}, nn^{\dag} \}\nonumber \\
& +2\phi_2^2\omega W_{\alpha} -i|\phi_3|^2[\p_{\alpha} nn^{\dag}-n\p_{\alpha} n^{\dag}+2(\p_{\alpha} n^{\dag}
n)\,nn^{\dag}] \nonumber \\
& -i(\phi_1-\phi_2)^2[\p_{\alpha}nn^{\dag}-n\p_{\alpha}n^{\dag}+2(\p_{\alpha} n^{\dag}
n)\,nn^{\dag})] \bigg) \nonumber \\
& +\frac{1}{r}\p_r(r\p_r)\gamma \tilde{W}_{\alpha}  -\frac{f^2}{r^2}\gamma[\{ \tilde{W}_{\alpha}, nn^{\dag}\}-2 nn^{\dag} \tilde{W}_{\alpha} nn^{\dag}] -\frac{g^2}{2}\bigg( (\phi_1^2+|\phi_3|^2-\phi_2^2)\gamma \{\tilde{W}_{\alpha}, nn^{\dag}
\}\nonumber \\
&+2\phi_2^2 \gamma \tilde{W}_{\alpha} -inn^{\dag}[(\p_{\alpha} \phi_3 )\phi_3^*- \phi_3\p_{\alpha} \phi_3^{*}-2|\phi_3|^2(\p_{\alpha} n^{\dag}
n)]  \bigg).
\end{align}
The first four lines are composed of  $W_{\alpha}$ terms, while the last two lines contain $\tilde{W}_{\alpha}$ terms. This equation is composed of
profiles, reducing matrices, and unknown matrices $W_{\alpha}$ and $\tilde{W}_{\alpha}$.

In order to solve it, we set $\tilde{N}=0$ at first. All $\tilde{W}_{\alpha}$ and $\phi_3$ terms disappear, since we return to the local vortex case. Then, the very remaining term can be separated into the profile part and orientational matrix part. In the spirit of the separation of variable method, all the orientational matrix parts should  have the same formula except a constant factor. Therefore, one has
\beq
W_{\alpha}=c_1 i[(\p_{\alpha}n)n^{\dag}-n\p_{\alpha}n^{\dag}+2(\p_{\alpha} n^{\dag} n) \,nn^{\dag}] =c_2 \{ W_{\alpha}, nn^{\dag}\}.
\eeq
One can check that $\Tr W_{\alpha}=0$, and $W_{\alpha}$ is Hermitian. The self consistent solution is $c_2=1$ and $c_1$ is undetermined. The term $nn^{\dag} W_{\alpha} nn^{\dag}$
vanishes in the equation of motion, since $nn^{\dag} W_{\alpha} nn^{\dag}=0$. In principle, one can always absorb the $c_1$ factor into the definition of $\omega$, and the final solution for the profile is unaffected. Therefore, we set $c_1=1$ without loss of generality, i.e.,
\beq
 W_{\alpha}=&i[(\p_{\alpha} n)n^{\dag}-n\p_{\alpha} n^{\dag}+2(\p_{\alpha} n^{\dag} n)nn^{\dag}].
  \eeq
Separating out the zero modes part, the equation of motion for $\omega(r)$ is written as follows
\beq \label{eq:omega}
0= \frac{1}{r}\p_r(r\p_r)\omega + \frac{f^2}{r^2}(1-\omega)
-\frac{g^2}{2}[  \omega (\phi_1^2+\phi_2^2+\phi_3^2)-(\phi_1-\phi_2)^2-
\phi_3^2 ].
 \eeq
With the help of BPS equations for profiles, the solution of $\omega(r)$ is given by
\beq \omega=1-\frac{\phi_1}{\sqrt{\xi}}. \eeq
This result has already been known in Ref. \cite{Gorsky:2004ad,Shifman:2004dr,Gudnason:2010rm,Shifman:2011xc}. This solution is the exact one,
which are independent of the coupling conditions.

Now let us consider the ``additional flavor'' part. Terms in the last two line of Eq.(\ref{eq:tem}) manifest properties of  semi local vortices.
The recipe of the solution for $\tilde{W}_{\alpha}$ is the same with that of $W_{\alpha}$. However, there is a caveat that $\phi_3$ contains the size modulus $\rho$, see Eq.(\ref{eq:phi3}). By substituting $\phi_3=(\rho /r^{k}) \phi_1$ into the last term  in Eq.(\ref{eq:tem}), one obtains
\beq
(\p_{\alpha} \phi_3 )\phi_3^*- \phi_3\p_{\alpha} \phi_3^{*}-2|\phi_3|^2(\p_{\alpha} n^{\dag}
n)= \frac{\phi_1^2}{r^{2k}}[\p_{\alpha} \rho \rho^* -\rho \p_{\alpha} \rho^* -2 |\rho|^2 (\p_{\alpha}n^{\dag} n)].
\eeq
Therefore, a natural solution for $\tilde{W}_{\alpha}$ is
\beq \label{eq:was}
\tilde{W}_{\alpha}= i   \left[ \p_{\alpha} \rho \rho^* -\rho \p_{\alpha} \rho^* -2 |\rho|^2 (\p_{\alpha}n^{\dag} n) \right]n n^{\dag}.
\eeq
Note that we have already assumed that the constant, namely $\tilde{c}_1 $, is set to be one, since it can be adjusted by redefining the profile $\gamma$. With this Ansatz, one easily finds that
\beq
 \{\tilde{W}_{\alpha}, nn^{\dag} \}=2\tilde{W}_{\alpha}, \quad  nn^{\dag}\tilde{W}_{\alpha} nn^{\dag}=\tilde{W}_{\alpha}.
\eeq
The equation of motion for $\gamma(r)$ is written as follows
\beq \label{eq:emgamma}
0= \frac{1}{r}\p_r(r\p_r)\gamma -\frac{g^2}{2}\bigg(
2(\phi_1^2+|\phi_3|^2)\gamma -\frac{\phi_1^2}{r^{2k}} \bigg).
\eeq
$\gamma$ has no exact analytical solution. In the strong coupling limit, the zeroth order solution of $\gamma$ is written as follows
\beq \gamma= \frac{1}{2\big(r^{2k}+|\rho|^2\big)}\ . \label{eq:gamma} \eeq
For $k=1$, this solution agrees with Shifman et al. \cite{Shifman:2011xc}. Now we have obtained the Ansatz for the gauge fields $A_{\alpha}$,
which indeed satisfies the equation of motion.

\subsection{The adjoint scalar}

 The Ansatz of the adjoint scalar has been less discussed than the gauge field $A_{\alpha}$. In the moduli matrix formalism, this Ansatz has been given in Ref. \cite{Eto:2011cv} \footnote{The Ansatz of the adjoint scalar in the field-theoretical method for the local non-Abelian vortices was given in the Ph.D thesis of the author.}. In the mass deformed theory, the effective potential  leads to  kink solutions on the vortex string. Recently, Bolokhov et al. proposed the Ansatz of the adjoint scalar for the local vortices \cite{Bolokhov:2013bea}. We will go beyond this, and give the Ansatz of the adjoint scalar for the semi local case.

 The procedure to obtain the Ansatz is the same as that of the gauge field. First, we give the equation of motion for the adjoint scalar, which
  is written as follows
\beq \label{eq:adjsca}
\frac{2}{g^2} \D_i \D^i \Phi+\{\Phi,q q^{\dag}\}+2  q {\bf M} q^{\dag}=0.
\eeq
The bulk matrix ${\bf M}$ has been given in Sec. \ref{sec:model}, and can be written as follows
\beq \label{eq:mass}
{\bf M}=\begin{pmatrix}
       {\bf M}_1 & \\
        & {\bf M}_2
        \end{pmatrix},
\eeq
where ${\bf M}_1=(m_1, \cdots, m_{N_c})$ and ${\bf M}_2=(m_{N_c+1}, \cdots, m_{N_f})$. In the vacuum, the adjoint scalar $\Phi$ takes the form
in Eq.(\ref{eq:vacuum}). The Ansatz of $\Phi$ describes that  fluctuations of zero modes excite in the background of vacuum.
Thus, we assume that the Ansatz of $\Phi$ is written as follows
 \beq \label{eq:phi2}
\Phi=-{\bf M}_1+(1- b(r)) \hat{\Phi}+ \chi(r)\Omega,
\eeq
where $\hat{\Phi}$ and $\Omega$ describe orientational modes and  size  modes, respectively. Note that we take the form of $(1-b(r))$  for
convenience in calculation.

Substituting $\Phi$ into the equation of motion in Eq.(\ref{eq:adjsca}), the equation of motion becomes
\begin{align}
0=&\left[ -(b''+\frac{b'}{r})-g^2 (1-b)\phi_2^2 \right]\hat{\Phi}+\left[ -(1-b)\frac{f^2}{r^2}-\frac{g^2}{2} (1-b)(\phi_1^2+\phi_3^2-\phi_2^2)\right] \{ nn^{\dag}, \hat{\Phi}\}\nonumber \\
& +\left[ \frac{f^2}{r^2}+\frac{g^2}{2}(\phi_1^2+\phi_3^2-\phi_2^2)-g^2 \phi_2 (\phi_1-\phi_2)\right]\{ nn^{\dag}, {\bf M}_1\} +2(1-b)\frac{f^2}{r^2} nn^{\dag} \hat{\Phi} nn^{\dag} \nonumber \\
 &-\left[ \frac{f^2}{r^2}+\frac{g^2}{2}(\phi_1-\phi_2)^2\right]2nn^{\dag}{\bf M}_1 nn^{\dag}+ \left( \chi'' + \frac{\chi'}{r} - g^2 \phi_2^2 \chi\right)\Omega \nonumber \\
 &+\left[-\chi \frac{f^2}{r^2}-\frac{g^2}{2} \chi (\phi_1^2 +\phi_3^2-\phi_2^2)\right]\{nn^{\dag},\Omega\} +2 \chi \frac{f^2}{r^2} nn^{\dag} \Omega nn^{\dag} -g^2  \phi_3^2 m_{N_c+1} nn^{\dag}.\label{eq:nnm}
 \end{align}
Since only one ``additional flavor'' was considered, the matrix ${\bf M}_2$ becomes $m_{N_c+1}$.

In order to solve $\Phi$, we set $m_{N_c+1}=0$ and $\Omega=0$ at first. In Eq.(\ref{eq:adjsca}), there are two evident terms related to the orientational moduli, i.e., $\{ nn^{\dag}, {\bf M}_1\}$ and $nn^{\dag}{\bf M}_1nn^{\dag}$. In the spirit of the separation of variables, the formula of $\hat{\Phi}$ should be a combination of them. Note that the coefficient of $\{ nn^{\dag}, {\bf M}_1\}$ and the coefficient of $2nn^{\dag}{\bf M}_1nn^{\dag}$ are similar except a $g^2 \phi_3^2/2$ coefficient. However, $\phi_3$ contains the size modulus $\rho$, thus one can  attribute it to the equation of motion for  $\chi$. In this way, a natural solution for $\hat{\Phi}$ is written as follows
\beq \label{eq:hatphisol}
\hat{\Phi}=\{ nn^{\dag},{\bf M}_1\}-2 nn^{\dag}{\bf M}_1nn^{\dag}.
\eeq
This formula also agrees with that given by Bolokhov et al., which is $\hat{\Phi}=\left[nn^{\dag},[nn^{\dag},{\bf M}_1]\right]$ \cite{Bolokhov:2013bea}. One can easily chech that such relations hold
\beq
\{nn^{\dag}, \hat{\Phi}\}=\hat{\Phi}, \qquad nn^{\dag}\hat{\Phi} nn^{\dag}=0.
\eeq
It is also remarkable that $\hat{\Phi}$ and $W_{\alpha}$ have the same relations with $nn^{\dag}$. With the help of these relations,
the equation of motion for the profile $b(r)$ is written as follows
\begin{align} \label{eq:sb}
0=&-b''-\frac{b'}{r}+b \frac{f^2}{r^2} -\frac{g^2}{2} \left[ (1-b)(\phi_1^2+\phi_3^2+\phi_2^2) -(\phi_1-\phi_2)^2-\phi_3^2 \right].
\end{align}
Compared with Eq.(\ref{eq:omega}), the solution of $b(r)$ is
\beq b=\frac{\phi_1}{\sqrt{\xi}}.\label{eq:b}
\eeq
The relation of $b$ and $\omega$ is $b+\omega=1$.

Now, let us throw away the settings, which are $m_{N_c+1}=0$ and $\Omega=0$. Remember that a term $g^2 \phi_3 nn^{\dag}{\bf M}_1nn^{\dag}$ has been
attributed to the rest of Eq.(\ref{eq:adjsca}). Therefore, the equation of motion for $\Omega$ is written as follows
\begin{align} \label{eq:chi}
0=& \left( \chi'' + \frac{\chi'}{r} - g^2 \phi_2^2 \chi\right)\Omega-\left[\chi \frac{f^2}{r^2}+\frac{g^2}{2} \chi (\phi_1^2 +\phi_3^2-\phi_2^2)\right]\{nn^{\dag},\Omega\}  \nonumber \\
&+2 \chi \frac{f^2}{r^2} nn^{\dag} \Omega nn^{\dag} +g^2 \phi_3^2 \big(nn^{\dag}{\bf M}_1 nn^{\dag}- m_{N_c+1} nn^{\dag}\big).
\end{align}
Following the same routine, the solution of $\Omega$ reads
\beq
\Omega=  nn^{\dag}{\bf M}_1 nn^{\dag}- m_{N_c+1} nn^{\dag}.
\eeq
With this Ansatz, the relations of $\Omega$ and $nn^{\dag}$ are
\beq  \label{eq:omerel} \{ nn^{\dag}, \Omega \}=2\Omega, \qquad nn^{\dag} \Omega nn^{\dag}=\Omega . \eeq
Note that  relations between $nn^{\dag}$ and $\Omega$ are the same with  relations between $nn^{\dag}$ and $\tilde{W}_{\alpha}$. Substituting them into Eq.(\ref{eq:chi}), one obtains the equation of motion
for $\chi$, i.e.,
\beq
0=\chi''+\frac{\chi'}{r}-g^2 \left[ \chi (\phi_1^2+\phi_3^3)-\phi_3^2 \right].
\eeq
In the strong coupling limit, $g\to \infty$, we have the solution for $\chi$, i.e.,
\beq
\chi=\frac{\phi_3^2}{\phi_1^2+\phi_3^2} =\frac{|\rho|^2}{r^{2k}+|\rho|^2}.
\eeq
Note that this solution respecting the exact relation of $\phi_3$ and $\phi_1$. Although there is no size moduli in the solution of $\Omega$, $\chi$ contains it. The solution of $\chi$ is similar to $\gamma$.
Now the complete Ansatze for $A_{\alpha}$ and $\Phi$ can be summarized as follow
\begin{align}
A_{\alpha}= &(1-\frac{\phi_1}{\sqrt{\xi}}) i[(\p_{\alpha}n)n^{\dag}-n\p_{\alpha}n^{\dag}+2(\p_{\alpha} n^{\dag} n) \,nn^{\dag}]  \nonumber \\
&+ \frac{i}{2\big(r^{2k}+|\rho|^2\big)} \big[ \p_{\alpha} \rho \rho^*-\rho \p_{\alpha} \rho^* -2 |\rho|^2 (\p_{\alpha}n^{\dag} n) \big]n n^{\dag}, \label{eq:Aalpha1}\\
\Phi =& -{\bf M}_1+(1- \frac{\phi_1}{\sqrt{\xi}})[\{ nn^{\dag},{\bf M}_1\}-2 nn^{\dag}{\bf M}_1nn^{\dag}] \nonumber \\
   &+\frac{|\rho|^2}{r^{2k}+|\rho|^2}[nn^{\dag}{\bf M}_1 nn^{\dag}- m_{N_c+1} nn^{\dag}]. \label{eq:phixx}
\end{align}
The size modulus $\rho$ couples with the orientational $n$. With these Ansatze, we are ready to construct the effective theory in the next step.

\section{The mass deformed effective action \label{sec:efa}}

The world-sheet action of the semi local non-Abelian vortices can be obtained by integrating out the vortex transverse plane, i.e.,
 \begin{align} \label{eq:eff1}
 S_{\rm eff}=& \int d^4x  {\cal L}^{2}_{\rm eff} \nonumber \\
           =&\int d^2x_{\alpha}  \int d^2x_i \Tr \bigg\{ \frac{1}{g^2} F_{i \alpha} F_i^{\alpha}+\D_{\alpha} q (\D^{\alpha}q)^{\dag}- \frac{1}{g^2}|\D_{i}\Phi|^2-|\Phi q+ q{\rm \bf M}|^2    \bigg\}.
 \end{align}
Note that we have changed the signs of relative terms according to the metric \footnote{The Minkowski metric is taken to be $(1,-1,-1,-1)$.}. With the vortex configuration (Eqs.(\ref{eq:q}) and (\ref{eq:Ai})) and  Ansatze (Eqs.(\ref{eq:Aalpha1}) and (\ref{eq:phixx})), one can perform the
integration over $d^2 x_i$ straight forward. The four terms in Eq.(\ref{eq:eff1}) above can be classified according to  properties of the effective action, which are the kinetic part and the  potential part. The kinetic effective action originates from the first and the second terms in Eq.(\ref{eq:eff1}), which are calculated to be
\begin{align}
{\cal L}^{2}_{\rm eff-kin}=&\Tr \left[ \frac{1}{g^2}  F_{i \alpha} F_i^{\alpha}+\D_{{\alpha}} q
  (\D^{\alpha}q)^{\dag} \right] \nonumber \\
  =&2 [\p_{\alpha}n^{\dag}\,\p^{\alpha} n + (\p_{\alpha} n^{\dag}
n)(\p^{\alpha} n^{\dag}n)]\bigg\{ \frac{1}{g^2}\big[\omega'^2+\frac{f^2}{r^2}(1-\omega)^2\big]
\nonumber \\
&+(1-\omega)(\phi_1-\phi_2)^2+\frac{\omega^2}{2}(\phi_1^2+\phi_2^2+|\phi_3|^2)-\omega|\phi_3|^2
\bigg\} \nonumber \\
&+[(\rho n)^{\dag}\p_{\alpha}(\rho n) -\p_{\alpha}(\rho n)^{\dag} \rho n ]^2
\bigg\{ -\frac{1}{g^2}\gamma'^2+ \gamma \frac{\phi_1^2}{r^{2k}}
-\gamma^2(\phi_1^2+\phi_3^2) \bigg\} \nonumber \\
&+\p_{\alpha}(\phi_3
n)^{\dag}\p^{\alpha}(\phi_3n)+(\p_{\alpha}\phi_1)(\p^{\alpha}\phi_1)+ \frac{1}{g^2}\frac{1}{r^2}(\p_{\alpha}f)(\p^{\alpha}f). \label{eq:effkin}
 \end{align}

In the  second last line, we have used an identity
\beq \label{eq:i1}
(\p_{\alpha}\rho )\rho^*- \rho\p_{\alpha}\rho^{*}-2|\rho|^2(\p_{\alpha} n^{\dag}n)=(\rho
n)^{\dag}\p_{\alpha}(\rho n) -\p_{\alpha}(\rho n)^{\dag} \rho n.
\eeq
We also use a square sign to replace the same formula  with a superscript $\alpha$ in the bracket for saving space. Some other tips are given in the appendix.
The existence of the last two terms seems to contradict the spirit of the separation of variable method, since we assume them to be independent of moduli
parameters in the beginning. In the BPS equations (Eq.(\ref{eq:bps1}) to (\ref{eq:bps4})), they are profiles without coupling to $\rho$. The size modulus enters in the exact relation in Eq.(\ref{eq:phi3}), and $\rho$ is considered to be only a complex number. However, when the effective theory is constructed, $\rho$ in profiles $\phi_1(r)$ and $f(r)$ should be considered as the size modulus. Thus, the derivative $
\p_{\alpha}$ of $\phi_1$, $\phi_3$ and $f$ are
not zero.

The integration over the transverse plane of the vortex can be realized by using the polar coordinates, i.e., $\int d^2 x_i \to 2\pi \int_0^{\infty} r dr $. The integration of profiles in the first two lines in Eq.(\ref{eq:effkin}) are well known \cite{Shifman:2004dr,Shifman:2006kd,Shifman:2011xc,Gudnason:2010rm}. Here, we repeat the exact result, which is written as follows
\begin{align}
\Gamma =& 2\pi \int^{\infty}_0 dr \, r\bigg[
\frac{1}{g^2}\big[\omega'^2+\frac{f^2}{r^2}(1-\omega)^2\big] +(1-\omega)(\phi_1-\phi_2)^2
\nonumber  \\
&+\frac{\omega^2}{2}(\phi_1^2+\phi_2^2+|\phi_3|^2)-\omega|\phi_3|^2
\bigg] =\frac{2\pi }{g^2}k \ .
\end{align}
The remaining lines in Eq.(\ref{eq:effkin}) is exact, but  profiles  $\phi_1$ and $f$ are known up to the order of $1/g^2$. Substituting all
the solutions of profiles, i.e. Eqs.(\ref{eq:phi3}), (\ref{eq:phi1}), (\ref{eq:phi1f}) and (\ref{eq:gamma}), into the last two lines of Eq.(\ref{eq:effkin}), one obtains
\begin{align} \label{eq:kinsem}
&[(\rho n)^{\dag}\p_{\alpha}(\rho n) -\p_{\alpha}(\rho n)^{\dag} \rho n ]^2
\bigg\{ -\frac{1}{g^2}\gamma'^2+ \gamma \frac{\phi_1^2}{r^{2k}}
-\gamma^2(\phi_1^2+\phi_3^2) \bigg\} \nonumber \\
&+\p_{\alpha}(\phi_3
n)^{\dag}\p^{\alpha}(\phi_3n)+(\p_{\alpha}\phi_1)(\p^{\alpha}\phi_1)+ \frac{1}{g^2}\frac{1}{r^2}(\p_{\alpha}f)(\p^{\alpha}f) \nonumber \\
=&[\p_{\alpha}n^{\dag}\,\p^{\alpha} n + (\p_{\alpha} n^{\dag}
n)^2]\;\frac{ \xi |\rho|^4}{(r^{2k}+|\rho|^2)^2} \big[1-\frac{4k}{g^2 \xi} \frac{r^{2k-2}(k r^{2k}+|\rho|^2)}{(r^{2k}+|\rho|^2)^2}\big] \, \nonumber \\
&+\, |\p_{\alpha}(\rho n)|^2 \;\frac{\xi\, r^{2k}}{(r^{2k}+|\rho|^2)^2}\big[1+\frac{4k(k-1)}{g^2 \xi} \frac{r^{2k-2}|\rho|^2}{(r^{2k}+|\rho|^2)^2}\big].
\end{align}
The first term  can be considered as a correction to the world-sheet action of the
local vortices. When the integration over $r$ is performed, there is no divergence in the first term. The second term, i.e. $|\p_{\alpha}(\rho n)|^2$ , is the kinematic term for the size modulus coupling to the orientational modes. This can be considered as a characteristic property of  semi local vortices. Except for
$k=1$, the result of the integration for the second term is also finite. Terms proportional to $1/g^2 \xi $ can be considered as the $1/g^2$ corrections.
The expression in Eq.(\ref{eq:kinsem}) seems to be different with Shifman et al. \cite{Shifman:2011xc}. This is not the fact, since we take the following decomposition, i.e.,
\beq \label{eq:decomposition}
|\p_{\alpha} (\rho n)^{\dag} \rho n |^2 = |\rho|^2 |\p_{\alpha} (\rho n)|^2-|\rho|^4[\p_{\alpha}n^{\dag}\p_{\alpha}n+(n^{\dag}\p_{\alpha}n)^2].
\eeq
The integration of the transverse plane can be done analytically. We give some examples for different values of  $k$ in the next section.

Similarly,  the effective potential from the last two terms in Eq.(\ref{eq:eff1}) is expressed as
\begin{align} \label{eq:effpot}
{\cal L}^2_{\rm eff-pot}=&-\frac{1}{g^2} \Tr(\D_i \Phi)^2 -\Tr |\Phi q +q {\bf M}|^2 \nonumber \\
=&-2 \Tr \left[{\bf M}_1^2 nn^{\dag}-{\bf M}_1nn^{\dag} {\bf M}_1 nn^{\dag} \right] \bigg\{
\frac{1}{g^2}\left( b'^2+ \frac{f^2}{r^2}b^2\right)  \nonumber \\
& +\frac{b^2-1}{2}(\phi_1^2+\phi_3^2+\phi_2^2) +2 \phi_1\phi_2 (1-b)+(\phi_1-\phi_2)^2 +\frac{\phi_3^2}{2}\bigg\} \nonumber \\
&-\Tr \left[ \left(nn^{\dag} {\bf M}_1 nn^{\dag}- m_{N_c+1} nn^{\dag}\right)^2 \right] \bigg\{ \frac{1}{g^2}\chi'^2
+\chi^2 (\phi_1^2+ \phi_3^2)-2\chi \phi_3^2 +\phi_3^2\bigg\}.
\end{align}
Except the last $\phi_3^2/2$ term in the second line, the first two lines describe the effective potential for the local vortex case.
After the integration (excluding the last $\phi_3^2/2$ term), we have such a formula, i.e.,
\beq \label{eq:em}
 -\frac{4\pi k}{g^2} \Tr \left[{\bf M}_1^2 nn^{\dag}-{\bf M}_1nn^{\dag} {\bf M}_1 nn^{\dag} \right] .
\eeq
In the formula of components, Eq.(\ref{eq:em}) agrees with Shifman et al. \cite{Shifman:2011xc}, except a total shift \footnote{The adjoint scalar is shifted in the Higgs branch.}. We attribute  the $\phi_3^2/2$ term to the contribution from the ``additional flavor''. In the same way, substituting the profiles into the remaining terms in Eq.(\ref{eq:effpot}), one obtains the following equation
\begin{align} \label{eq:potsem}
&- \Tr \left[{\bf M}_1^2 nn^{\dag}-{\bf M}_1nn^{\dag} {\bf M}_1 nn^{\dag} \right] \phi_3^2  \nonumber \\
&-\Tr \left[ \left(nn^{\dag} {\bf M}_1 nn^{\dag}- m_{N_c+1} nn^{\dag}\right)^2 \right] \bigg\{ \frac{1}{g^2}\chi'^2
+\chi^2 (\phi_1^2+ \phi_3^2)-2\chi \phi_3^2 +\phi_3^2\bigg\}. \nonumber \\
=&-\Tr \left[{\bf M}_1^2 nn^{\dag}-{\bf M}_1nn^{\dag} {\bf M}_1 nn^{\dag} \right] \frac{|\rho|^2\xi}{r^{2k}+|\rho|^2}
\big( 1-\frac{4k}{g^2 \xi} \frac{|\rho|^2 r^{2k-2}}{(r^{2k}+|\rho|^2)^2} \big) \nonumber \\
&-\Tr \left[ \left(nn^{\dag} {\bf M}_1 nn^{\dag}- m_{N_c+1} nn^{\dag}\right)^2 \right] \frac{|\rho|^2\xi r^{2k}}{(r^{2k}+|\rho|^2)^2}\big[1+
\frac{4(k^2-k)}{g^2\xi} \frac{|\rho|^2 r^{2k-2}}{(r^{2k}+|\rho|^2)^2} \big].
\end{align}
We can see that the coefficients in last lines of both Eq.(\ref{eq:kinsem}) and Eq.(\ref{eq:potsem}) is exactly equal except the $|\rho|^2$ term.
However, the coefficients of first terms in these two equations are not the same. Can we take use some identity, similar to that in Eq.(\ref{eq:decomposition}), and change the formula in Eq.(\ref{eq:potsem})? Yes, we can. The identity reads
\begin{align} \label{eq:mi}
\Tr[({\bf M}_1-m_{N_c+1} \mathbbm{1}_{N_c})^2 nn^{\dag}]=& \Tr \left[{\bf M}_1^2 nn^{\dag}-{\bf M}_1nn^{\dag} {\bf M}_1 nn^{\dag} \right] \nonumber \\
 &+\Tr \left[ \left(nn^{\dag} {\bf M}_1 nn^{\dag}- m_{N_c+1} nn^{\dag}\right)^2 \right].
\end{align}
Substitute it into Eq.(\ref{eq:potsem}), one obtains the following
\begin{align} \label{eq:potsem1}
&- \Tr \left[{\bf M}_1^2 nn^{\dag}-{\bf M}_1nn^{\dag} {\bf M}_1 nn^{\dag} \right] \phi_3^2  \nonumber \\
&-\Tr \left[ \left(nn^{\dag} {\bf M}_1 nn^{\dag}- m_{N_c+1} nn^{\dag}\right)^2 \right] \bigg\{ \frac{1}{g^2}\chi'^2
+\chi^2 (\phi_1^2+ \phi_3^2)-2\chi \phi_3^2 +\phi_3^2\bigg\}. \nonumber \\
=&-\Tr \left[{\bf M}_1^2 nn^{\dag}-{\bf M}_1nn^{\dag} {\bf M}_1 nn^{\dag} \right] \frac{ |\rho|^4\xi }{(r^{2k}+|\rho|^2)^2} \big[1-\frac{4k}{g^2 \xi} \frac{r^{2k-2}(k r^{2k}+|\rho|^2)}{(r^{2k}+|\rho|^2)^2}\big]  \nonumber \\
&-\Tr[({\bf M}_1-m_{N_c+1} \mathbbm{1}_N)^2 (\rho n) (\rho n)^{\dag}] \frac{\xi r^{2k}}{(r^{2k}+|\rho|^2)^2}\big[1+
\frac{4(k^2-k)}{g^2\xi} \frac{|\rho|^2 r^{2k-2}}{(r^{2k}+|\rho|^2)^2} \big].
\end{align}
 We observe that the coefficients of two kinetic terms coincide with the coefficients of two potential terms, respectively. This is
 an interesting result, since the same result also happens for local vortices, where the coefficient is $4\pi k/g^2$. This also reduces a lot
 of calculation in the following.

By collecting all these pieces of work, the effective action reads
\begin{align} \label{eq:epf}
{\cal L}_{\rm eff}= & \left[ [\p_{\alpha}n^{\dag}\,\p^{\alpha} n + (\p_{\alpha} n^{\dag}n)(\p^{\alpha} n^{\dag}n)]- \Tr \big({\bf M}_1^2 nn^{\dag}-{\bf M}_1nn^{\dag} {\bf M}_1 nn^{\dag} \big) \right]  \nonumber \\
& \cdot \bigg\{ \frac{4k \pi}{g^2} + 2\pi \int rdr \big[ \frac{ |\rho|^4\xi }{(r^{2k}+|\rho|^2)^2} -\frac{4k}{g^2 } \frac{|\rho|^4 r^{2k-2}(k r^{2k}+|\rho|^2)}{(r^{2k}+|\rho|^2)^4}\big] \bigg\} \nonumber \\
&+\left[ \p_{\alpha}(\rho n)^{\dag} \p^{\alpha}(\rho n)- \Tr \big[ ({\bf M}_1-m_{N_c+1} \mathbbm{1}_N)^2 (\rho n) (\rho n)^{\dag}\big] \right] \nonumber \\
& \cdot \bigg\{ 2\pi \int r dr \big[ \frac{\xi r^{2k}}{(r^{2k}+|\rho|^2)^2}+
\frac{4(k^2-k)}{g^2} \frac{|\rho|^2 r^{4k-2}}{(r^{2k}+|\rho|^2)^4} \big] \bigg\}.
\end{align}
This formula reduces to the effective theory of local non-Abelian vortices, if we simply set $\rho=0$. The remaining job is integrating out
$r$ for given $k$.

\subsection{The fundamental  case}

First, we consider the $k=1$ case, which is fundamental. Since there is divergence, we introduce an infrared cut-off $L$ as the upper limit for integrating $r$ to tame the divergence. We also assume that $L$ is much larger than the size of the semi local vortex, i.e., $L \gg |\rho|$. The effective action is written as follows
\begin{align} \label{eq:ke1}
{\cal L}^{k=1}_{ \rm eff}=&\left(\frac{2\pi}{g^2}+ \pi \xi |\rho|^2\right) \bigg\{ [\p_{\alpha}n^{\dag}\,\p^{\alpha} n + (\p_{\alpha} n^{\dag}
n)(\p^{\alpha} n^{\dag}n)]-\Tr \left[{\bf M}_1^2 nn^{\dag}-{\bf M}_1nn^{\dag} {\bf M}_1 nn^{\dag} \right] \bigg\}  \nonumber \\
&+\left[ \pi \xi \left(2\log|\frac{L}{\rho}|-1\right)\right] \bigg\{ \p_{\alpha}(\rho n)^{\dag} \p^{\alpha} (\rho n) -\Tr \left[ \left({\bf M}_1- m_{N_c+1} \mathbbm{1}_{N_c}\right)^2 (\rho n)(\rho n)^{\dag}\right] \bigg\}.
\end{align}

The kinetic term agrees exactly with the result of Shifman et al.\footnote{One can compare the results by using the identity in Eq.(\ref{eq:decomposition}).}, see Eq.(2.49) in \cite{Shifman:2011xc}.
There is no divergence in the first term. The effective kinetic part is exactly the formula of the non-Abelian local vortex case. But, the
coefficient $4\pi/g^2$ was replaced by $2\pi/g^2+\pi \xi |\rho|^2$.  Note that we also considered the contributions from $\delta \phi_1$. This is remarkable, since the finite term of semi-local vortices seems to contain corrections to the effective action of local vortices. This kinetic term also
shows that the size modulus couples to orientational zero modes, this agrees with the prediction of the K\"ahler potential \cite{Eto:2007yv}.

The coefficient of the second line is divergent. This means that we need infinite energy to excite the motion of the size moduli. The finite energy requires that the size modulus $\rho$ is not dynamical, which must be fixed in principle. Then, one can extract the size modulus from the derivative $\p_{\alpha}$. Thus, the effective theory contains powers of $\rho \p_{\alpha}$. This agrees with the spirit of the effective dynamics \cite{Eto:2007yv}.
 The effective potential in Eq.(\ref{eq:ke1}) does not agree exactly with Shifman et al.,
since they did not construct the Ansatz for the adjoint scalar. However, ignoring the divergent coefficients, the two mass terms in our formula
match with the two mass terms in their reference.

The theory in Eq.(\ref{eq:ke1}) is exact up to the order of $1/g^2 \xi$.
We have four physical parameters in the effective theory, which are $L$, $|\rho|$, $1/g\sqrt{\xi} $, and $\delta m_i= m_i -m_{N_c+1}$ $(i=1, \dots, N_c)$, respectively. Their rations is important to construct the effective theory. First, we consider the $1/g\sqrt{\xi}$ parameter to be a constant, since
it is the size of the local non-Abelian vortices. The power expansion in Eq.(\ref{eq:expansion}) is valid, if the size of the semi local vortices $|\rho|$ is much larger than the size of the local vortices $1/g\sqrt{\xi}$, i.e., $|\rho|\gg 1/g\sqrt{\xi} $. The derivation from Eq.(\ref{eq:epf}) to Eq.(\ref{eq:ke1}) throws away some finite terms, which consider the limit $L \gg |\rho|$.
Remember that we also take the assumption that $m \ll \sqrt{\xi}$ in Section {\ref{sec:model}}.  The inverse of
$\delta m_i$ may be comparable to $L$. For example, Shifman et al. took $L$ to be $1/\delta m$ to rewrite the effective potential.

The lump limit can also be realized in the strong coupling limit $g \to \infty$, or equally by keeping $1/g\sqrt{\xi}$ finite, while taking $|\rho|$ to be infinite. Respecting $L \gg |\rho|$, the effective Lagrangian can
be further reduced to the following,
\beq \label{eq:lump}
{\cal L}^{k=1}_{ \rm eff}=2 \pi \xi |\rho|^2\log|\frac{L}{\rho}| \bigg\{ \p_{\alpha}n^{\dag} \p^{\alpha}  n - \sum_{i=1}^{N_c} \delta m_i^2 |n_i|^2\bigg\}.
\eeq
By introducing  a new parameter $z\equiv \rho (2\pi \xi \ln |L/ \rho|)^{1/2}$, the model in Eq.(\ref{eq:lump}) reduces to  the $zn$ model of Shifman et al..
However, if we don't consider the strong coupling limit, the theory in Eq.(\ref{eq:ke1}) does not coincide with Eq.(4.8) in Ref.\cite{Shifman:2011xc}.
The difference is only the finite integral, i.e., we have the coefficient $(\pi \xi|\rho|^2 +2\pi/g^2)$ in the first line of Eq.(\ref{eq:ke1}); while they used $4\pi/g^2$ in their paper.

The generalization to $\tilde{N}>1$ case is done straight forward. The effective action reads
\begin{align} \label{eq:ke1t}
{\cal L}^{k=1}_{ \rm eff}=&\left(\frac{4\pi}{g^2}+ \sum_{j=1}^{\tilde{N}}(\pi \xi |\rho_j|^2-\frac{2\pi}{g^2})\right) \bigg\{ [\p_{\alpha}n^{\dag}\,\p^{\alpha} n + (\p_{\alpha} n^{\dag}
n)(\p^{\alpha} n^{\dag}n)]-\big( \sum_{i=1}^{N_c} m_i^2 |n_i|^2 \nonumber \\
&-| \sum_{i=1}^{N_c}m_i |n_i|^2|^2  \big) \bigg\} + \sum_{j=1}^{\tilde{N}} \pi \xi R_j^2 \bigg\{ \p_{\alpha} (\rho_jn)^{\dag} \p^{\alpha} (\rho_j n ) -\sum_{i=1}^{N_c}(m_i- m_j)^2 |\rho_j n_i|^2 \bigg\}.
\end{align}
where $R_j \equiv [2\ln |L/ \rho_j|-1]^{1/2}$ is a real number. One can also shift a total (or averaged) mass to the mass matrix ${\bf M}_1$ without changing the effective potential \cite{Shifman:2011xc}.
The semi-local model has similarities in both the non-Abelian Higgs model and the sigma model lump. It may interpolate between them.

It is also interesting to construct the two dimensional ${\cal N}=(2,2)$  gauge theory from the effective action of vortices. First we study the local vortices case. The effective theory of local vortices is simply expressed as follows
\beq \label{eq:local}
{\cal L}^{k=1}_{\rm local} = \frac{4\pi}{g^2} \bigg\{ [\p_{\alpha}n^{\dag}\,\p^{\alpha} n + (\p_{\alpha} n^{\dag}
n) (\p^{\alpha} n^{\dag}n)]-\big( \sum_{i=1}^{N_c} m_i^2 |n_i|^2 -| \sum_{i=1}^{N_c}m_i |n_i|^2|^2  \big) \bigg\}
\eeq

 The component fields $n_i$
($i=1,\dots, N_c$) are considered as  chiral matter fields in the two dieminonal supersymmetric $U(1)$ gauge theory. The $U(1)$ vector multiplet contains the gauge fields $A_{\alpha}$ and a complex scalar $\sigma$. They can be constructed as follows
\beq  \label{eq:alpsig}
A_{\alpha}=in^{\dag}\p_{\alpha} n, \qquad \sigma= \sum_{i=1}^{N_c} m_i |n_i|^2.
\eeq
We also define the covariant derivative in the two dimensional theory to be ${\cal D}_{\alpha} \equiv \p_{\alpha}+i A_{\alpha}$.
Then, following Shifman et al., we rescale $n_i$ to get a D-term condition, which is written as follows
\beq
0=\sum_{i=1}^{N_c} |n_i|^2-\frac{4\pi}{g^2}.
\eeq
The effective theory of the local vortex in Eq.(\ref{eq:local}) becomes the following
\beq
{\cal L}^{k=1}_{\rm local}= \sum_{i=1}^{N_c} \left( |{\cal D}_{\alpha}n_i|^2 -|\sigma-m_i|^2|n_i|^2\right)
-\frac{e^2}{2} \big(\sum_{i=1}^{N_c} |n_i|^2-\frac{4\pi}{g^2} \big).
\eeq
By adding kinetic terms for both $A_{\alpha}$ and $\sigma$, the theory in Eq.(\ref{eq:local}) agrees with the Hanany-Tong model for the local vortices \cite{Hanany:2004ea}.
The theory has kink solutions which are confined monopoles on the vortex string \cite{Hanany:2004ea,Eto:2011cv,Shifman:2011xc}.

Now we study the semi-local vortex case. First, we observe that
\beq
 \p_{\alpha}(\rho n)^{\dag} \p^{\alpha} (\rho n) = (\tilde{\cal D}_{\alpha} \rho)^* \tilde{\cal D}^{\alpha} \rho,
  \eeq
where the covariant derivative is $\tilde{\cal D}_{\alpha}\equiv \p_{\alpha}-i A_{\alpha}$; while $A_{\alpha}$ is given in Eq.(\ref{eq:alpsig}). So $\rho$ can be considered as a chiral matter field with $U(1)$ charge $-1$. In the effective theory in Eq.(\ref{eq:ke1t}), coefficients proportional to $1/g^2$ disappear in the strong coupling limit. In order to match coefficients of  remaining terms, we set $\log |L/\rho|=1$. This means the integral radius is about ten times larger than the size of semi-local vortices, allowed by the physical condition. Besides, we define $\tilde{n}_j \equiv  \rho_j/ |\rho|$ ($j=1,\dots, \tilde{N}$), where  $|\rho|^2 =\sum_{j=1}^{\tilde{N}}|\rho_j|^2$  is the total size of the semi-local string. With these set-up,
the effective theory of semi local vortices in Eq.(\ref{eq:ke1t}) becomes
\begin{align} \label{eq:dual}
{\cal L}^{k=1}_{\rm sem-loc}=&\sum_{i=1}^{N_c}\left[ |{\cal D}_{\alpha} n_i|^2-2|\sigma-m_i|^2|n_i|^2\right]+\sum_{j=1}^{\tilde{N}}\left[ |\tilde{\cal D}_{\alpha} \tilde{n}_j|^2 - |\sigma-m_j|^2|\tilde{n}_j|^2\right].
\end{align}
There is a remarkable Seiberg-like duality in the theory. One can also construct a dual theory for the four dimensional $U(\tilde{N}_c)\times SU(N_f)$ supersymmetric gauge theory, by replacing $n$ and $\tilde{n}$ with each other in Eq.(\ref{eq:dual}) and defining $\sigma=\sum_{i=1}^{\tilde{N}}m_i |n_i|^2$. In Ref.\cite{Eto:2007yv}, the Seiberg-like duality was also manifested in the K\"ahler quotient construction. In our deduction, we take  that $L \sim 10 |\rho|$ and sizes of the semi-local vortices are normalized. These two conditions are very reasonable in the physical content. The theory in Eq.(\ref{eq:dual}) agrees with the Hanany-Tong construction of the semi local vortices \cite{Hanany:2004ea}, where the anti chiral matter fields $\tilde{n}_j$ with mass $m_j$ are identified with the normalized size moduli.

\subsection{The high winding case}
For the $k=2$ case, all terms become finite after the integration, if $L \gg |\rho|$. The effective Lagrangian is written as follows
\begin{align} \label{eq:ke2}
{\cal L}^{k=2}_{ \rm eff}=&\left(\frac{16\pi}{3g^2}+ \frac{\pi^2}{4} \xi |\rho|\right) \bigg\{ [\p_{\alpha}n^{\dag}\,\p^{\alpha} n + (\p_{\alpha} n^{\dag}
n)(\p^{\alpha} n^{\dag}n)]-\big( \sum_{i=1}^{N_c} m_i^2 |n_i|^2 -| \sum_{i=1}^{N_c}m_i |n_i|^2|^2  \big) \bigg\}  \nonumber \\
&+\left[ \frac{\pi^2}{4}\xi|\rho|^2+\frac{2\pi}{3g^2}\right] \bigg\{ \p_{\alpha}n^{\dag} \p^{\alpha}  n - \sum_{i=1}^{N_c} \delta m_i^2 |n_i|^2\bigg\}.
\end{align}
From the index theory, we know that the dimension of the $k=2$ moduli is $4N_f$. However, the Anzatz of $q$ contains  $n$ as the orientational  modes and $\rho$ as the size modulus, which is incomplete to denote all the moduli parameters. A more precise representation of the moduli parameters is given
in the moduli matrix method \cite{Eto:2007yv}. With the present formula, we need to understand where we stand on, and what is missing.

The squark field for $k=2$ semi-local vortices in the muduli matrix method is written as follows \cite{Eto:2007yv}
\beq \label{eq:qk2}
q(z)= \begin{pmatrix}
             z^2 - \alpha z -\beta &  & & & {\bf s}z + {\bf u}\\
              a_1 z +b_1 & \quad 1& & &-a_1 {\bf s}\\
             \vdots &&\ddots& &\vdots\\
            a_{N_c-1}z+b_{N_c-1}&&&1 & \quad -a_{N_c-1}{\bf s}
          \end{pmatrix},
\eeq
where ${\bf s} \equiv(s_1, \dots, s_{\tilde{N}})$ is a raw vector, and ${\bf u}$  has the similar definition. By the K\"ahler quotient construction,
the moduli parameters can be extracted from $q(z)$ to a set of matrices $\big( \mathbf{Z},\mathbf{\Psi},\tilde{\mathbf{\Psi}}\big)$.
The divergent part of the K\"ahler potential can be written as follows
\beq  \label{eq:kahler}
{\cal K}= 2\pi \xi \log L \Tr|\mathbf{\Psi}\tilde{\mathbf{\Psi}}| ^2 .
\eeq
In the case of the configuration in Eq.(\ref{eq:qk2}), the three matrices are expressed as
\beq \label{eq:3matrices}
 \mathbf{Z}=\begin{pmatrix}
            0 &  1\\
             \beta & \alpha
                       \end{pmatrix}, \quad
 \mathbf{\Psi}  =\begin{pmatrix}
            1&  0\\
            b_1 & a_1\\
            \vdots & \vdots \\
            b_{N_c-1} & a_{N_c-1}
                       \end{pmatrix},  \quad
 \tilde{\mathbf{\Psi}}=\begin{pmatrix}
           {\bf s}\\
           \alpha {\bf s} + {\bf u}
                       \end{pmatrix},
\eeq
respectively. It is not difficult to obtain that
\begin{align}
{\cal K}= 2\pi \xi \log L& \bigg\{ \sum_j^{\tilde{N}}|s_j|^2\big(1+|\alpha|^2 \sum_{i=1}^{N_c-1}|a_i|^2 +\sum_{i=1}^{N_c-1}( \alpha b_i^* a_i +c.c.)\big)  \nonumber \\
&+(\sum_j^{\tilde{N}}|{u}_j|^2) (\sum_{i=1}^{N_c-1}|a_i|^2)+\big[(\sum_j^{\tilde{N}}u_j s^*_j) \cdot \sum_{i=1}^{N_c-1}(b_i^* a_i + \alpha^* |a_i|^2) + c.c.\big] \bigg\}.
\end{align}
When we transform the vortex configuration in Eq.(\ref{eq:q}) into the moduli matrix form, while keeping the same moduli parameters, we find that $\alpha=0$, $a_i=0$, and ${\bf s}=0$ in Eq.(\ref{eq:qk2}). We can see that the divergent part of the K\"ahler potential disappears completely. $\alpha=0$ means that the positions of two fundamental semi-local vortices coincide. $a_i =0$  denotes that $N_c-1$  orientational parameters are set to zero. From ${\bf s}=0$, we know that only  half of size moduli were considered.  Therefore, the Lagrangian in Eq.(\ref{eq:ke2}) describes a reduced point of the muduli space.

Now a question arises, i.e.,  how to construct the effective theory for  high winding semi-local vortices. Following our recipe, we find that it is a difficult task. We do not know how to represent all the modulus parameters in $q$. For example, in the $k=2$ case, the profile in the additional flavor part should be written as $\phi_3e^{i \theta} +\phi_4 $. One can find reduced  BPS equations, if $\phi_3=\frac{\rho_1}{r} \phi_1$ and $\rho_4=\frac{\rho_4}{r^2} \phi_1$. However, this will bring the crossing terms like $\rho_3^* \rho_4 e^{i \theta}$ in the BPS equations. The following
calculations will explode. We prefer to choose the moduli matrix method to construct the high winding semi-local vortices in a generic point of the moduli space.

\section{Conclusion and discussion}

In this paper, we have constructed the mass deformed effective theory of the semi local non-Abelian vortices. The Ansatze for the adjoint scalar and the gauge field were solved via  EL equations, this ensures the minimal energy excitation of zero modes. The separation of variable method is powerful in
the calculation of EL equations. We obtained all solutions of profiles up to the $1/g^2$ order. The Ansatze for the adjoint scalar and  gauge fields have
the similar structure, respecting the vortex configuration. The effective theory of the semi local vortices were constructed by integrating out the transverse plane of the vortex string. It was found that the size modulus couples to  orientational zero modes, and the kinetics of the size modulus  is a divergent term. The effective theory interpolates between the local vortices and the sigma model lump. The relation between our mass deformed effective theory and the $zn$ model was clarified. We also showed that the effective theory has a Seiberg-like duality respecting certain normalization conditions.
The mass deformed theory of high winding semi local vortices was studied. There are no divergent terms in the effective theory, because the vortex configuration denotes only a special point in the moduli space. We leave the construction of the high winding semi-local vortices as a future work.

The field theoretical method to construct the effective world sheet theory of non-Abelian vortices offers a special proof of Dorey's 2d-4d duality \cite{Dorey:1999zk}. Our work extends the verification of the 2d-4d duality to the $N_f > N_c$ case. Although we did not discuss the classical spectrum of our effective action, it has no difference with Shifman et al \cite{Shifman:2011xc}, and agrees with that of the D-brane construction \cite{Hanany:2004ea}. The 2d-4d duality sheds light on the connection between  supersymmetry gauge theories  and quantum integrable systems  \cite{Dorey:2011pa,Chen:2011sj}, which has been studied by Nekrasov and Shatashvili \cite{Nekrasov:2009rc,Nekrasov:2009uh,Nekrasov:2009ui}. For example, the two dimensional twisted superpotential is identified  the Yang-Yang function of the integrable system; while the supersymmetric vacua can be identified with Bethe states of quantum integrable systems. As illustrated in the present paper, the twisted superpotential can be derived from non-Abelian vortices. This gives us a hint that there is a dictionary between non-Abelian vortices and integrable systems. Let us check the route from vortices to integrable systems. First we give a vortex configuration and construct its mass deformed effective theory. From the effective theory, one obtains the classical vacuum, i.e., a kink solution corresponding a non-Abelian monopole. This can be considered as a Bethe state of quantum integrable systems. However, the moduli space of non-Abelian vortices are very rich \cite{Eto:2011cv}. For examples, $\mathbbm{C}P^{N-1}$ and $Gr(k,N)$ represent 
orientational zero modes of $U(N)$ vortices. The representation of the moduli space of non-Abelian vortices with winding number $k$ looks like Bethe states of  spin chains, see Section 2 in Ref.\cite{Eto:2011cv}. Following this route, one can find more integrable systems from the vortex section. This is quite non-trivial, since very limited integrable systems are known.  The present work brings us one more question, i.e., how the mass deformed potential of semi local vortices is connected with integrable systems. All these questions are interesting future research topics.

\acknowledgments

The author thanks Sven Bjarke Gudnason and Jarah Evslin for useful discussions. This work had been supported  by the China Postdoctoral Science Foundation funded Project with grant No. 2012M510548.

\bigskip
\bigskip
\appendix{\bf Calculation tips}

We have used many tips to reduce the calculation, especially the traces. We list them here for the convenience of readers.
The traces used in the calculation of the kinetic effective action are written as follows
\begin{align}
 \Tr(W_{\alpha})^2&=2\Sigma, &\qquad  \Sigma &\equiv  [(\p_{\alpha}n^{\dag}n)^2+\p_{\alpha}n^{\dag}\p_{\alpha}n],  \\
\Tr[W_{\alpha}\tilde{W}_{\alpha}]&=0, &\qquad 0&=\Tr[\p_{\alpha}(nn^{\dag}) \,nn^{\dag}], \\
\Tr[W_{\alpha}^2 nn^{\dag}]&=\Sigma, &\qquad 2\Sigma&=\Tr [\big(\p_{\alpha}(nn^{\dag})\big)^2],\\
 \Tr[\p_{\alpha}(nn^{\dag})W_{\alpha}]&=0, &\qquad 0&=\Tr[nn^{\dag}W_{\alpha}], \\
\Tr[\p_{\alpha}(nn^{\dag})W_{\alpha}nn^{\dag}]&=i\Sigma. &\qquad&
\end{align}
The traces used for the calculation of the effective potential are given in the following
\begin{align}
\Tr[\hat{\Phi}^2]&=2\Xi, &\qquad  \Xi &\equiv \Tr[{\rm M}_1^2 nn^{\dag}-{\rm M}_1 nn^{\dag}{\rm M}_1 nn^{\dag}] \\
\Tr[\Omega^2]&=\Theta,  &\qquad \Theta &\equiv \Tr[nn^{\dag}{\rm M}_1nn^{\dag} -m_{N_c+1} nn^{\dag}]^2, \\
\Tr [ \Omega{\rm M}_1]&= \Lambda, &\qquad \Lambda &\equiv \Tr [{\rm M}_1nn^{\dag}{\rm M}_1 nn^{\dag}-m_{N_c+1} {\rm M}_1 nn^{\dag}] \\
\Tr[ \Omega nn^{\dag}]&= \Delta , &\qquad \Delta &\equiv \Tr [{\rm M}_1 nn^{\dag} -m_{N_c+1} ],\\
 \Tr[\hat{\Phi}^2nn^{\dag}]&=\Xi,  &\qquad 2 \Xi&=\Tr[\hat{\Phi}{\rm M}_1],\\
\Tr[\{nn^{\dag},{\rm M}_1\}\hat{\Phi}]&=2 \Xi,& \qquad 0&=\Tr[\hat{\Phi} \Omega],  \\
\Tr[\Omega^2 nn^{\dag}]&=\Theta, &\qquad 2 \Lambda&=\Tr[\{nn^{\dag},{\rm M}_1\}\Omega], \\
\Tr[{\rm M}_1 \hat{\Phi}]&=2 \Xi,  &\qquad  0&=\Tr[nn^{\dag}\hat{\Phi}].
\end{align}


\begin{thebibliography}{99}

\bibitem{Auzzi:2003fs}
  R.~Auzzi, S.~Bolognesi, J.~Evslin, K.~Konishi and A.~Yung,
  ``NonAbelian superconductors: Vortices and confinement in N=2 SQCD,''
  Nucl.\ Phys.\ B {\bf 673}, 187 (2003)
  [hep-th/0307287].

\bibitem{Hanany:2003hp}
  A.~Hanany and D.~Tong,
  ``Vortices, instantons and branes,''
  JHEP {\bf 0307}, 037 (2003)
  [hep-th/0306150].

\bibitem{Shifman:2004dr}
  M.~Shifman and A.~Yung,
  ``NonAbelian string junctions as confined monopoles,''
  Phys.\ Rev.\ D {\bf 70}, 045004 (2004)
  [hep-th/0403149].

  \bibitem{Hanany:2004ea}
  A.~Hanany, D.~Tong,
  ``Vortex strings and four-dimensional gauge dynamics,''
  JHEP {\bf 0404 } (2004)  066.
  [hep-th/0403158].

\bibitem{Shifman:2006kd}
  M.~Shifman and A.~Yung,
  ``Non-Abelian semilocal strings in N=2 supersymmetric QCD,''
  Phys.\ Rev.\ D {\bf 73}, 125012 (2006)
  [hep-th/0603134].

\bibitem{Gorsky:2004ad}
  A.~Gorsky, M.~Shifman and A.~Yung,
  ``Non-Abelian meissner effect in Yang-Mills theories at weak coupling,''
  Phys.\ Rev.\ D {\bf 71}, 045010 (2005)
  [hep-th/0412082].



\bibitem{Shifman:2011xc}
  M.~Shifman, W.~Vinci and A.~Yung,
  ``Effective World-Sheet Theory for Non-Abelian Semilocal Strings in N = 2 Supersymmetric QCD,''
  Phys.\ Rev.\ D {\bf 83}, 125017 (2011)
  [arXiv:1104.2077 [hep-th]].

\bibitem{Bolokhov:2013bea}
  P.~A.~Bolokhov, M.~Shifman and A.~Yung,
  ``Twisted-Mass Potential on the Non-Abelian String World Sheet Induced by Bulk Masses,''
  Phys.\ Rev.\ D {\bf 88}, 085016 (2013)
  [arXiv:1308.4494 [hep-th]].



\bibitem{Cipriani:2012pa}
  M.~Cipriani and T.~Fujimori,
  ``Effective Action of Non-Abelian Monopole-Vortex Complex,''
  arXiv:1207.2070 [hep-th].

\bibitem{Eto:2007yv}
  M.~Eto, J.~Evslin, K.~Konishi, G.~Marmorini, M.~Nitta, K.~Ohashi, W.~Vinci, N.~Yokoi,
  ``On the moduli space of semilocal strings and lumps,''
  Phys.\ Rev.\  {\bf D76 } (2007)  105002.
  [arXiv:0704.2218 [hep-th]].

\bibitem{Gudnason:2010rm}
  S.~B.~Gudnason, Y.~Jiang, K.~Konishi,
  ``Non-Abelian vortex dynamics: Effective world-sheet action,''
  JHEP {\bf 1008 } (2010)  012.
  [arXiv:1007.2116 [hep-th]].

\bibitem{Eto:2011cv}
  M.~Eto, T.~Fujimori, S.~B.~Gudnason, Y.~Jiang, K.~Konishi, M.~Nitta and K.~Ohashi,
  ``Vortices and Monopoles in Mass-deformed SO and USp Gauge Theories,''
  JHEP {\bf 1112}, 017 (2011)
  [arXiv:1108.6124 [hep-th]].
\bibitem{Dorey:1999zk}
  N.~Dorey, T.~J.~Hollowood and D.~Tong,
  ``The BPS spectra of gauge theories in two-dimensions and four-dimensions,''
  JHEP {\bf 9905}, 006 (1999)
  [hep-th/9902134].


\bibitem{Eto:2012qda}
  M.~Eto, T.~Fujimori, M.~Nitta, K.~Ohashi and N.~Sakai,
  ``Higher Derivative Corrections to Non-Abelian Vortex Effective Theory,''
  Prog.\ Theor.\ Phys.\  {\bf 128}, 67 (2012)
  [arXiv:1204.0773 [hep-th]].

\bibitem{Eto:2006cx}
  M.~Eto, K.~Konishi, G.~Marmorini, M.~Nitta, K.~Ohashi, W.~Vinci and N.~Yokoi,
  ``Non-Abelian Vortices of Higher Winding Numbers,''
  Phys.\ Rev.\ D {\bf 74}, 065021 (2006)
  [hep-th/0607070].


\bibitem{Delduc:1984sz}
F.~Delduc and G.~Valent,
``Classical And Quantum Structure Of The Compact Kahlerian Sigma Models,''
Nucl.\ Phys.\  B {\bf 253}, 494 (1985);
F.~Delduc and G.~Valent,
``Renormalizability Of The Generalized Sigma Models Defined On Compact
Hermitian Symmetric Spaces,''
Phys.\ Lett.\  B {\bf 148} (1984) 124.


\bibitem{Vachaspati:1991dz}
  T.~Vachaspati and A.~Achucarro,
  ``Semilocal cosmic strings,''
  Phys.\ Rev.\ D {\bf 44}, 3067 (1991).

\bibitem{Hindmarsh:1991jq}
  M.~Hindmarsh,
  ``Existence and stability of semilocal strings,''
  Phys.\ Rev.\ Lett.\  {\bf 68}, 1263 (1992).


  \bibitem{Leese:1992fn}
  R.~A.~Leese and T.~M.~Samols,
  ``Interaction of semilocal vortices,''
  Nucl.\ Phys.\ B {\bf 396}, 639 (1993).

\bibitem{Manton:1981mp}
  N.~S.~Manton,
  ``A Remark on the Scattering of BPS Monopoles,''
  Phys.\ Lett.\ B {\bf 110}, 54 (1982).


\bibitem{Eto:2006uw}
  M.~Eto, Y.~Isozumi, M.~Nitta, K.~Ohashi and N.~Sakai,
  ``Manifestly supersymmetric effective Lagrangians on BPS solitons,''
  Phys.\ Rev.\ D {\bf 73}, 125008 (2006)
  [hep-th/0602289].

\bibitem{Sakai:2007fm}
  N.~Sakai, M.~Eto, Y.~Isozumi, M.~Nitta and K.~Ohashi,
  ``Effective Lagrangians on Domain Walls and Other Solitons,''
  PoS STRINGSLHC {\bf }, 025 (2006)
  [hep-th/0703136 [HEP-TH]].


\bibitem{Eto:2006pg}
  M.~Eto, Y.~Isozumi, M.~Nitta, K.~Ohashi and N.~Sakai,
  ``Solitons in the Higgs phase: The Moduli matrix approach,''
  J.\ Phys.\ A {\bf 39}, R315 (2006)
  [hep-th/0602170].


\bibitem{Dorey:2011pa}
  N.~Dorey, S.~Lee and T.~J.~Hollowood,
  ``Quantization of Integrable Systems and a 2d/4d Duality,''
  JHEP {\bf 1110}, 077 (2011)
  [arXiv:1103.5726 [hep-th]].

\bibitem{Chen:2011sj}
  H.~-Y.~Chen, N.~Dorey, T.~J.~Hollowood and S.~Lee,
  ``A New 2d/4d Duality via Integrability,''
  JHEP {\bf 1109}, 040 (2011)
  [arXiv:1104.3021 [hep-th]].


\bibitem{Nekrasov:2009uh}
  N.~A.~Nekrasov and S.~L.~Shatashvili,
  ``Supersymmetric vacua and Bethe ansatz,''
  Nucl.\ Phys.\ Proc.\ Suppl.\  {\bf 192-193}, 91 (2009)
  [arXiv:0901.4744 [hep-th]].

\bibitem{Nekrasov:2009ui}
  N.~A.~Nekrasov and S.~L.~Shatashvili,
  ``Quantum integrability and supersymmetric vacua,''
  Prog.\ Theor.\ Phys.\ Suppl.\  {\bf 177}, 105 (2009)
  [arXiv:0901.4748 [hep-th]].
  
\bibitem{Nekrasov:2009rc}
  N.~A.~Nekrasov and S.~L.~Shatashvili,
  ``Quantization of Integrable Systems and Four Dimensional Gauge Theories,''
  arXiv:0908.4052 [hep-th].





\end{thebibliography}
\end{document}